\documentclass[
reprint,
superscriptaddress,
amsmath,
amssymb,
aps,
pra
]{revtex4-2}
\usepackage{graphicx}
\usepackage{dcolumn}
\usepackage{bm}
\usepackage[
bookmarks=true, 
colorlinks=true, 
linkcolor=blue, 
urlcolor=blue, 
citecolor=blue
]{hyperref}
\usepackage{graphicx}
\usepackage{xcolor}
\usepackage{tikz}
\usepackage{color}
\usepackage[normalem]{ulem}
\usepackage{enumitem}

\begin{document}

\preprint{APS/123-QED}

\title{Diffusion-Enhanced Optimization of Variational Quantum Eigensolver for General Hamiltonians}

\author{Shikun Zhang}
\thanks{These authors contributed equally to this work}
\affiliation{School of Physics, Beihang University, Beijing 100191, China}

\author{Zheng Qin}
\thanks{These authors contributed equally to this work}
\affiliation{School of Physics, Beihang University, Beijing 100191, China}

\author{Yongyou Zhang}
\affiliation{School of Physics, Beijing Institute of Technology, Beijing 100081, China}

\author{Yang Zhou}
\email[Contact author: ]{yangzhou9103@buaa.edu.cn}
\affiliation{School of Physics, Beihang University, Beijing 100191, China}
    
\author{Rui Li}
\affiliation{School of Physics, Beihang University, Beijing 100191, China}

\author{Chunxiao Du}
\affiliation{School of Physics, Beihang University, Beijing 100191, China}

\author{Zhisong Xiao}
\affiliation{School of Physics, Beihang University, Beijing 100191, China}
\affiliation{School of Instrument Science and Opto-Electronics Engineering, Beijing Information Science and Technology University, Beijing 100192, China}

\begin{abstract}

Variational quantum algorithms (VQAs) have emerged as a promising approach for achieving quantum advantage on current noisy intermediate-scale quantum devices. However, their large-scale applications are significantly hindered by optimization challenges, such as the barren plateau (BP) phenomenon, local minima, and numerous iteration demands. In this work, we leverage denoising diffusion models (DMs) to address these difficulties. The DM is trained on a few data points in the Heisenberg model parameter space and then can be guided to generate high-performance parameters for parameterized quantum circuits (PQCs) in variational quantum eigensolver (VQE) tasks for general Hamiltonians. Numerical experiments demonstrate that DM-parameterized VQE can explore the ground-state energies of Heisenberg models with parameters not included in the training dataset. Even when applied to previously unseen Hamiltonians, such as the Ising and Hubbard models, it can generate the appropriate initial state to achieve rapid convergence and mitigate the BP and local minima problems. These results highlight the effectiveness of our proposed method in improving optimization efficiency for general Hamiltonians.

\end{abstract}	

\maketitle 

\section{Introduction}

In the current noisy intermediate-scale quantum (NISQ) era \cite{NISQ1, NISQ2, NISQ3, NISQ4}, variational quantum algorithms (VQAs) \cite{VQA1} have emerged as a promising approach for near-term practical applications, such as finding ground energy, combinatorial optimization, and quantum machine learning \cite{2-10-ent, 2-11-ent, 2-12-ent, 2-15-ent, 2-17-ent, 2-19-ent}.
VQAs employ a hybrid quantum-classical framework. A parameterized quantum circuit (PQC) generates target quantum states, and a classical computer implements parameter optimization guided by information extracted from measurements on these states. 
The classical optimization process has twofold importance: enabling the continuous adjustment of quantum gate parameters to explore solution space and mitigating the impact of noisy qubits and operational imperfections, while maintaining low quantum resource demands \cite{1}.

Despite its importance, classical optimization has become the bottleneck that limits the power of quantum computation. On the one hand, the gradients of the cost function vanish exponentially with system size \cite{EXPvsBP-28, EXPvsBP-29, 7-27-ent, EXPvsBP-31, EXPvsBP-32, yuanxiao-199}, requiring an exponential number of measurement shots to resolve and determine the cost-minimizing direction. This is just the notorious barren plateau (BP) phenomenon \cite{bp-original}. While various strategies have been proposed to mitigate BP \cite{2-28, 2-29, 2-30, 2-31, 2-32, 2-33, 2-34, 2-35}, the associated exponential complexity continues to undermine the quantum advantage promised by VQAs \cite{VQA1, bp-original}. Furthermore, the non-convex energy landscape of VQAs, which contains numerous local minima \cite{2-24, 2-25}, exacerbates optimization challenges by easily trapping the training process.
On the other hand, exploring solution space often requires numerous iterations, leading to substantial measurement shot demands \cite{2-26, 2-27}. Moreover, the training process is typically more susceptible to noise and decoherence than the inference process, making optimization costly and necessitating high-fidelity quantum devices and extensive measurement resources \cite{2}.

The variational quantum eigensolver (VQE), as the first proposed VQA, is aimed to investigate various properties of a series of parameterized Hamiltonians, including the ground-state energy as a function of Hamiltonian parameters, the energy gap between the ground and excited states, the convergence behavior of specific parameterized circuits, and the evolution of observables under external influences \cite{1, 2-4-ent, 2-5-ent, 2-6-ent, 2-7-ent, 2-8-ent, 2-9-ent, qinzheng, yuanxiao-198}. 
In standard VQE setups, the optimization process must be repeated many times for each Hamiltonian with a specific parameter setting, amplifying the aforementioned optimization challenges. 
Several previous studies have suggested incorporating Hamiltonian encoding into the VQE framework to handle general Hamiltonians \cite{1,2}, yet few have approached this issue from an optimization perspective.

\begin{figure*}
\centering
\includegraphics[width=1\textwidth]{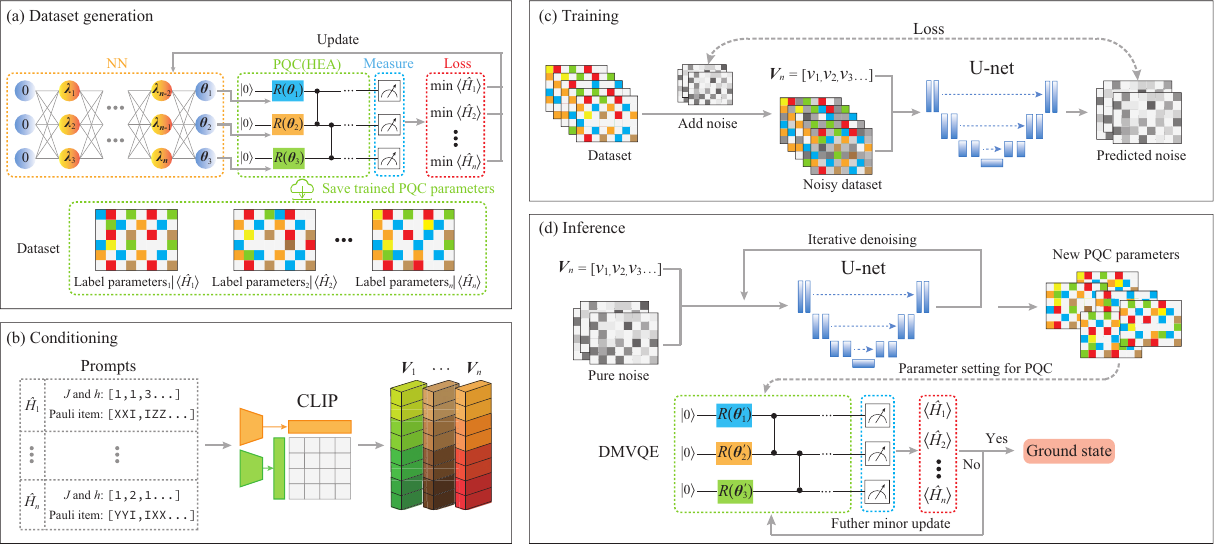}
  \caption{\textbf{PQC parameter generation framework.} (a) Using the method proposed in Ref.\cite{NNVQE} to obtain the label parameters corresponding to specific Hamiltonian. (b) Creation of the diffusion model’s conditioning. Prompts of general Hamiltonians are transformed into continuous vectors through a pre-trained CLIP encoder. (c)-(d) Schematic representation of the training of the diffusion model and the posterior inference from the trained model. See text for details.}
 \label{fig1}
\end{figure*}

In this work, we address the problems stated above by utilizing denoising diffusion models (DMs) \cite{5-18}, a state-of-the-art generative machine learning approach \cite{5-19}. This approach applies to general local interaction Hamiltonians of interest to physicists which can be expressed in the form
\begin{eqnarray}
  \hat{H} = \sum_{i} c_i P_i,
  \label{eq1}
\end{eqnarray}
where $c_i$ are coefficients and $P_i$ are tensor products of Pauli operators ${I, X, Y, Z}$. Specifically, the DM is trained with prompts describing a certain representative Hamiltonian to iteratively denoise corrupted samples (label parameters) within the training dataset, enabling the acquisition of high-performance PQC parameters for general Hamiltonians. Once trained, the DM generates new parameters that are distinct yet closely resemble the label parameters, which can be subsequently utilized to configure PQC in VQE.
It should be pointed out that the PQC parameter setting plays a crucial role in the optimization process, as it ensures the algorithm starts in a region of the Hilbert space where a solution is likely to exist \cite{1}. VQE can only converge to the correct solution if an appropriate initial state or suitable parameters for the PQCs are provided. 
In contrast, random parameter settings are more susceptible to BP and local minima. Our results demonstrate that DM-parameterized VQE (DMVQE) significantly accelerates convergence and mitigates the impact of BP and local minima, thereby enhancing optimization efficiency compared to randomly parameterized VQE (RPVQE). More importantly, DMVQE exhibits remarkable generalization capability. Guided by prompts, it can seamlessly extend to Hamiltonian parameter regimes beyond those of the dataset and even generalize to entirely different Hamiltonian structures without any fine-tuning, highlighting its adaptability to general Hamiltonians.

Additionally, we found that even assigning DM-generated parameters to only a portion of deeper PQCs can help mitigate the BP problem.
It is known that a sufficiently deep PQC is typically required to solve most problems of interest. However, deep PQCs with high expressibility and entanglement are more susceptible to the BP \cite{EXPvsBP}.
Our study shows that configuring DM-generated parameters to the first few layers of a deeper PQC while randomly initializing the remaining layers, termed DMVQE$'$, can significantly mitigate the BP problem.
We validate the effectiveness of our approach by applying it to the Heisenberg, Ising, and Hubbard models, demonstrating its potential to address real-world quantum computational problems.

\section{Methods}

The PQC parameters are represented in a data structure similar to image data, making it natural to treat them as images within the DM framework. In this context, the reverse process of the DM can be viewed as the transition from randomly initialized parameters (standard normal noise) to optimal parameters (clear images). By iterative denoising the randomly initialized parameters using the DM, this process mimics the optimization procedure in VQE, facilitating the generation of high-performance PQC parameters. This section comprehensively describes the proposed framework, including dataset generation, diffusion model architecture, and generative performance.

\subsection{Dataset generation}

 To train the DM, label parameters (i.e., PQC parameters that enable VQE to achieve high performance) corresponding to specific Hamiltonians must first be obtained. The dataset generation process includes the following three elements.

\emph{Structure of PQCs---} The construction of PQC, or ansatz, typically depends on the specific tasks it is intended for. In this work, we adopt the hardware-efficient ansatz (HEA), which is designed with gates that can be directly implemented on near-term quantum hardware. A typical HEA unit layer consists of single-qubit gates with tunable parameters and two-qubit gates to provide the necessary entanglement. Generally, HEA requires the unit layer to be repeated $L$ times to improve their computational performance \cite{HEA}. This can be represented as follows,
\begin{eqnarray}
  U(\boldsymbol\theta) =\prod_{l=1}^{L}U_{l}(\theta_{l})W_{l},
  \label{equ2}
\end{eqnarray}
with 
\begin{eqnarray}
  U_{l}(\theta_{l}) = \bigotimes_{j=1}^{N}R_{\sigma}(\theta_{l}^{j}),
  \label{equ3}
\end{eqnarray}
where $R_{\sigma}(\theta_{l}^{j}) = e^{-i\theta_{l}^{j}\sigma/2}$ with $\sigma \in (\sigma_{x},\sigma_{y},\sigma_{z})$ being one of the Pauli matrices. $l$ represents a specific layer of HEA, whose total number is $L$. $j$ denotes the $j$th qubit with the total number being $N$. $W_{l}$ are unparametrized two-qubit gates (e.g., $CZ$ gates). Without loss of generality, we set the total layers $L$ to be 10.

\emph{Hamiltonian for training---} To make the DM have the best generalization ability, the training dataset should be constructed with a representative Hamiltonian $\hat{H}$. In this work, we utilize the one-dimensional Heisenberg model with 8-qubit to construct the dataset. This is motivated by two key factors: First, as a canonical model of local interactions, it captures couplings between the $x$-, $y$-, and $z$-components of spin operators on neighboring lattice sites, enabling a Pauli decomposition spanning $I$, $X$, $Y$, $Z$; Second, its ground-state energies can be computed with high accuracy using VQE, enabling the extraction of high-performance PQC parameters.

The Heisenberg model has the following Hamiltonian,
\begin{eqnarray}
  \hat{H} = J\sum_{j=1}^{N}(S_{j}^{x}S_{j+1}^{x} + S_{j}^{y}S_{j+1}^{y} + S_{j}^{z}S_{j+1}^{z}) + h\sum_{j=1}^{N}S_{j}^{z}
  \label{eq4}
\end{eqnarray}
where $S_j$ represents the spin of the $j$th electron, with the coupling strength $J\in [1,4]$ and the magnetic field strength $h\in [1,4]$. $J$ and $h$ are discretized with a step size of 0.1, constructing 961 Hamiltonian samples based on all possible combinations.

\emph{Obtaining label parameters---} For each Hamiltonian sample, we apply the method introduced in Ref. \cite{NNVQE}, referred to as NNVQE, to obtain the label parameters.
Specifically, the outputs of a neural network are used to parameterize the PQC within the standard VQE framework, as illustrated in Fig.~\ref{fig1}(a). This is critical to the generative performance of the DM. DMs inherently belong to the domain of supervised learning, and the quality of the label parameters directly impacts the DM's generative performance. NNVQE helps mitigate the BP problem to some extent \cite{NNVQE}, facilitating the acquisition of high-performance PQC parameters. Given the complexity of the VQA optimization landscape \cite{4-16,4-25}, we conduct repeated experiments to identify the parameters closest to the optimal solution. The parameters corresponding to the lowest energy, referred to as the label parameters, are then stored to construct the dataset for the subsequent training process.

\subsection{Diffusion models architecture}

DMs rooted in non-equilibrium thermodynamics \cite{4-38,4-39}, are generative models designed to progressively remove noise from input data to generate clear data. The denoising diffusion probabilistic model \cite{4-32} is a key implementation of DMs. It includes three main steps for PQC parameter generation.

\emph{Conditioning---} We first construct the conditioning to guide the DM training and generation, which, in our study, corresponds to the Hamiltonian information, as illustrated in Fig.~\ref{fig1}(b). Each $c_i$ and $P_i$ in Eq.~(\ref{eq1}) encompass information from all axes and thus can be regarded as prompts describing the Hamiltonian. These prompts are transformed into continuous vectors using a pre-trained Contrastive Language–Image Pretraining (CLIP) encoder \cite{CLIP}. The CLIP encodes each $c_i$ and $P_i$ as a 512-dimensional vector. The final Hamiltonian representation is derived by averaging these vectors to provide a unified representation for various Hamiltonians.

\emph{Training---} We employ a training framework consisting of forward and reverse processes, each performed over multiple time steps $t$, as shown in Fig.~\ref{fig1}(c). In the forward process, Gaussian noise is incrementally added to the training samples (label parameters) $X_0$ and obtain the noisy samples $X_t$, 
\begin{eqnarray}
  X_t = \sqrt{\bar{\alpha}_t}X_0 + \sqrt{1-\bar{\alpha}_t}\boldsymbol\epsilon_t,
\end{eqnarray}
where $\boldsymbol\epsilon_t \sim \mathcal{N}(0, \boldsymbol{I})$ denotes Gaussian noise. $\bar{\alpha}_t$ is a variance schedule that decreases as $t$ increases, defined as $\bar{\alpha}_t=\prod_{i=0}^{t}\alpha_i=\prod_{i=0}^{t}(1-\beta_i)$, where $\beta_i$ is a constant less than 1, and $\beta_0$ to $\beta_t$ increase linearly. The DM is trained to estimate each corrupted sample's noise and denoise it progressively in the reverse process via the noise prediction network, ultimately reconstructing $X_0$. The formal expression is as follows,
\begin{eqnarray}
  X_{t-1} \sim \frac{1}{\sqrt{\alpha_t}}(X_t - \frac{1-\alpha_t}{\sqrt{1-\bar{\alpha}_t}}\boldsymbol\epsilon_\theta(X_t, t)),
\end{eqnarray}
where $\boldsymbol\epsilon_\theta(X_t, t)$ is the noise predicted by the noise prediction network, and $\bar{\alpha}_t=\prod_{i=t}^{0}\alpha_i$. The U-net architecture \cite{5-19} is employed as the noise prediction model which features an encoder-decoder design with skip connections and incorporates residual convolutional layers \cite{5-29}.

\emph{Inference---} Upon completion of training, a pure noise is first randomly sampled, and the trained DM iteratively denoises it based on the given guiding conditions, ultimately producing a high-performance PQC parameter after several steps, as illustrated in the upper part of Fig.~\ref{fig1}(d). The randomness of the pure noise affects the DM's generative performance. To mitigate this effect, we repeat the generation process 100 times for each Hamiltonian and select the parameters corresponding to the lowest ground-state energy in a single run as the DM-generated parameters. 

We refer to the aforementioned working mechanism as DM-parameterized VQE (DMVQE), as illustrated in the lower part of Fig.~\ref{fig1}(d).
Ideally, the DM-generated parameters can be directly configured into the PQC to obtain the ground-state energy of the corresponding Hamiltonian in a single run, eliminating the need for an expensive optimization process. Considering the discrepancy between the label parameters in the dataset and the optimal parameters, as well as the inherent performance limitations of the DM, most of the DM-generated parameters do not meet the desired accuracy. In such cases, only a small number of optimization epochs are required. 

\subsection{Generative performance}

\begin{figure}
\centering
\includegraphics[width=0.48\textwidth]{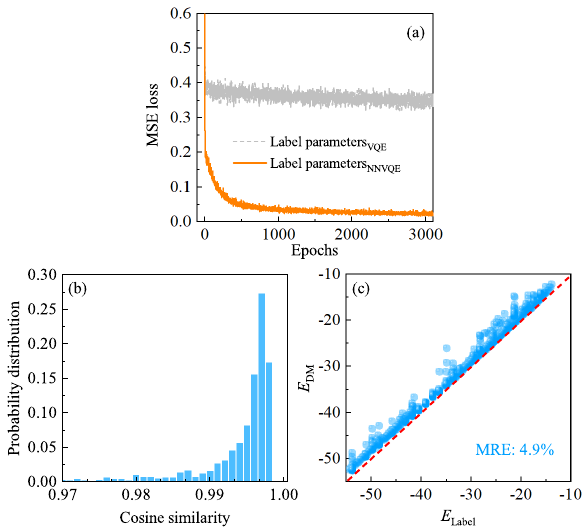}
  \caption{\textbf{Generative performance.} (a) Learning curves during the U-net training for different label parameter acquisition methods. (b) The probability distribution of the cosine similarity between the DM-generated parameters and the label parameters. (c) Confusion scatter plot based on the ground-state energies corresponding to the label parameters ($E_{\rm Label}$) and those obtained from the DM-generated parameters ($E_{\rm DM}$) in a single run.}
  \label{fig2}
\end{figure}

We begin by emphasizing the necessity of building a dataset with NNVQE. When training the U-net with label parameters obtained from standard VQE, it exhibits poor learnability, with the loss function prematurely converging to a high value with significant fluctuations, indicating ineffective optimization, as shown by the gray curve Fig.~\ref{fig2}(a). In contrast, training with label parameters obtained from NNVQE demonstrates superior learnability, with the loss function steadily decreasing and converging to a lower value, reflecting successful optimization, as illustrated by the orange curve in Fig.~\ref{fig2}(a). This superiority can be attributed to two main reasons. First, label parameters generated by the RPVQE are in the range of $-2\pi$ to $2\pi$, whereas NNVQE constrains parameter values to the range of -1 to 1 \cite{NNVQE}, significantly reducing dataset complexity. Second, the joint training of the neural network and PQCs in NNVQE results in the label parameters with more distinctive features, while parameters from RPVQE resemble random noise, hindering effective denoising.

Next, we verify the generative performance of the DMs from two aspects. The cosine similarity between the DM-generated parameters (viewed as vector $\boldsymbol{A}$) and the label parameters (viewed as vector $\boldsymbol{B}$) was calculated using the following formula
\begin{eqnarray}
cosine\ similarity = cos(\theta) = \frac{\boldsymbol{A}\cdot{\boldsymbol{B}}}{\|\boldsymbol{A}\|\|\boldsymbol{B}\|}
\label{eq7}
 \end{eqnarray}
to evaluate the quality of the DM-generated parameters. Fig.~\ref{fig2}(b) illustrates the distribution of cosine similarity. It can be seen that the majority of cosine similarity values exceed 0.99, with none equal to 1, indicating that the DM-generated parameters are distinct from but highly similar to the label parameters. Besides, confusion scatter plots are plotted based on the ground-state energies corresponding to the label parameters ($E_{\rm Label}$) and those obtained from the DM-generated parameters ($E_{\rm DM}$) in a single run, as shown in Fig.~\ref{fig2}(c). The horizontal axis represents $E_{\rm Label}$, while the vertical axis represents $E_{\rm DM}$, with the dashed regression line indicating ideal alignment. The confusion scatter plot shows a strong consistency between $E_{\rm Label}$ and $E_{\rm DM}$. Notably, some points lie directly on the regression line, indicating that $E_{\rm DM}$ is nearly equal to their true corresponding values. This suggests that the DM-generated parameters can even directly yield highly accurate ground state energies for some Hamiltonians in the dataset. The mean relative error (MRE) of 4.9\% between $E_{\rm DM}$ and $E_{\rm Label}$ further confirms the DM's generative capability.

\section{Results}

In this section, we test the proposed DMVQE for general Hamiltonians with three examples: the Heisenberg, Ising, and Hubbard models. Notably, the same pre-trained DM model was used in all cases without any additional fine-tuning, highlighting its adaptability to general Hamiltonians. Numerical experiments demonstrate its effectiveness from three perspectives: 
(\romannumeral1) it can explore the ground-state energies of the Heisenberg model, even when its parameters are not included in the training dataset; (\romannumeral2) it can explore the ground-state energies of previously unseen Ising models, achieving rapid convergence and alleviating the local minima problem; (\romannumeral3) it can generate initial states for Hubbard models with entirely different Hamiltonian structures, enabling rapid convergence and mitigating the BP problem.

\subsection{The Heisenberg model}

\begin{figure}
\centering
\includegraphics[width=0.48\textwidth]{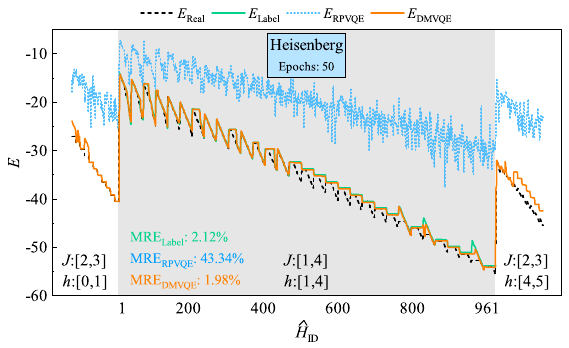}
  \caption{\textbf{Performance of DMVQE for Heisenberg model.} Comparison of ground state energy ($E$) distributions obtained from various computational approaches, organized by Hamiltonian IDs for the Heisenberg model.}
  \label{fig3}
\end{figure}

Based on the pre-trained DM, we first evaluate the performance of DMVQE in calculating the ground-state energies for the Heisenberg model. The real ground-state energies $E_{\rm Real}$, computed with exact diagonalization, serve as the gold standard baseline. We also compare our results $E_{\rm DMVQE}$ with the $E_{\rm Label}$ obtained by NNVQE and $E_{\rm RPVQE}$ obtained by RPVQE.

The shaded region in Fig.~\ref{fig3} illustrates the ground-state energy distribution of 961 Hamiltonians after 50 optimization steps. It is shown that $E_{\rm DMVQE}$ surpasses $E_{\rm Label}$ across the dataset. Specifically, on the left of the shaded area, $E_{\rm DMVQE}$ closely aligns with $E_{\rm Real}$, whereas slight deviations are observed on the right of the shaded area, likely due to imperfections in the label parameters. Overall, the MRE of $E_{\rm DMVQE}$ compared to $E_{\rm Real}$ is 1.98\%, outperforming the 2.12\% MRE of $E_{\rm Label}$. Meanwhile, $E_{\rm RPVQE}$ exhibit significant discrepancies from $E_{\rm Real}$, with an MRE of 43.34\%. The results indicate that even for Hamiltonian parameter regimes beyond those of the dataset, DMVQE can effectively accelerate the convergence to real ground energy.

More importantly, DMVQE demonstrates exceptional generalization capability. As shown in Fig.~\ref{fig3}, the non-shaded regions on both sides represent the ground-state energy distribution of 242 Hamiltonians with $J$ and $h$ outside the dataset. In the left non-shaded region, $E_{\rm DMVQE}$ is almost perfectly aligned with $E_{\rm Real}$, with an MRE of 1.96\%, achieving high accuracy. In the right non-shaded region, slight deviations in $E_{\rm DMVQE}$ and $E_{\rm Real}$ are observed, resulting in an MRE of 7.73\%. By contrast, $E_{\rm RPVQE}$ exhibits substantial deviations from $E_{\rm Real}$ in both non-shaded regions, with MREs of 44.15\% and 43.18\%, respectively. The results indicate that DMVQE effectively explores the ground-state energies of the Heisenberg model, even when the Hamiltonian parameters are not included in the training dataset.

\subsection{The Ising model}

\begin{figure}
\centering
\includegraphics[width=0.49\textwidth]{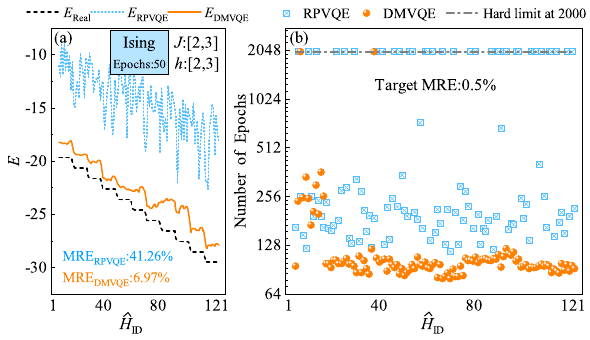}
  \caption{\textbf{Performance of DMVQE for Ising model.} (a) Comparison of ground state energy ($E$) distributions obtained from various computational approaches, organized by Hamiltonian IDs for the Ising model. (b) The number of epochs required to achieve a 0.5\% target MRE for the Ising model. A hard limit of 2000 epochs is set to prevent convergence to local minima leading to infinite loops.}
  \label{fig4}
\end{figure}

The Ising model is a typical quantum many-body model with wide applications in both fundamental physics and quantum algorithms (e.g., Quantum Approximate Optimization Algorithm). Its Hamiltonian can be written as
\begin{eqnarray}
  \hat{H} = J\sum_{\langle{i,j}\rangle}(S_{i}^{z}S_{j}^{z}) + h\sum_{i}S_{i}^{z}
  \label{eq8}
\end{eqnarray}
where $S_i$ denotes the spin of the $i$th electron, $J$ represents the coupling strength, and $h$ specifies the magnetic field strength. To further evaluate the adaptability of DMVQE to general Hamiltonians, we employ DMVQE to compute the ground-state energies of the Ising model with $J\in [2,3]$ and $h\in [2,3]$.

Fig.\ref{fig4}(a) illustrates the ground-state energy distribution of 121 Hamiltonians for the Ising model. To be compared with Fig.\ref{fig3}, the number of optimization epochs is also set to 50. After 50 epochs, the performance of $E_\mathrm{DMVQE}$ significantly surpasses that of $E_\mathrm{RPVQE}$, showing a trend and proximity much closer to $E_\mathrm{Real}$. Specifically, compared to $E_\mathrm{Real}$, the MRE of $E_\mathrm{DMVQE}$ is 6.79\%, while that of $E_\mathrm{RPVQE}$ reaches only 41.26\%. This demonstrates that for unseen Hamiltonians, DMVQE can generate an initial state much closer to the exact solution and thus accelerate the optimization process.

To directly illustrate its advantage of accelerating convergence, we set the target MRE to 0.5\% and compared the number of epochs required by DMVQE and RPVQE to reach this target. A hard limit of 2000 epochs is imposed to prevent optimization from stagnating at local minima and failing to achieve the target MRE, resulting in an infinite loop. As shown in Fig.\ref{fig4}(b), for 121 Hamiltonians of the Ising model, the number of epochs required by DMVQE to achieve the target MRE is significantly lower than that of RPVQE in most cases. DMVQE encountered local minima in only two Hamiltonians. For the remaining Hamiltonians, DMVQE successfully reached the target MRE with fewer optimization epochs. In contrast, RPVQE not only required more epochs to achieve the target MRE but also failed for as many as 35 Hamiltonians, which were trapped in local minima. This strongly demonstrates that DMVQE can not only accelerate convergence but also effectively mitigate the impact of local minima.

\subsection{The Hubbard model}

\begin{figure}
\centering
\includegraphics[width=0.42\textwidth]{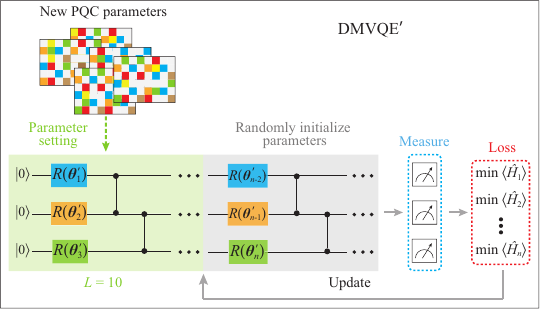}
  \caption{\textbf{Schematic diagram of DMVQE$'$.} Specifically, the DM-generated parameters are assigned to the first few layers of a deeper PQC (the first 10 layers in our work), while the remaining layers are randomly initialized, followed by global optimization.}
  \label{fig5}
\end{figure}

The Hubbard model is a key framework in condensed matter physics for studying strongly correlated electron systems. Its Hamiltonian is given by
\begin{eqnarray}
  \hat{H} = -T\sum_{\langle{i,j} \rangle,\sigma}(\hat{c}_{i,\sigma}^{\dagger}\hat{c}_{j,\sigma}+\hat{c}_{j,\sigma}^{\dagger}\hat{c}_{i,\sigma}) + U\sum_{i}\hat{n}_{i\uparrow}\hat{n}_{i\downarrow},
\end{eqnarray}
where $T$ represents electron hopping between sites $\langle{i,j} \rangle$, $U$ is the on-site coulomb repulsion, $\hat{c}_{i,\sigma}^{\dagger}(\hat{c}_{i,\sigma})$ are creation (annihilation) operators, and $n_{i\sigma}=\hat{c}_{i,\sigma}^{\dagger}\hat{c}_{i,\sigma}$ is the number operator. Through the Jordan-Wigner transformation, the Hubbard model can be expressed in the form of Eq.~(\ref{eq1}), with a Hamiltonian structure that is entirely different from and unseen in the dataset.

\begin{figure}
\centering
\includegraphics[width=0.49\textwidth]{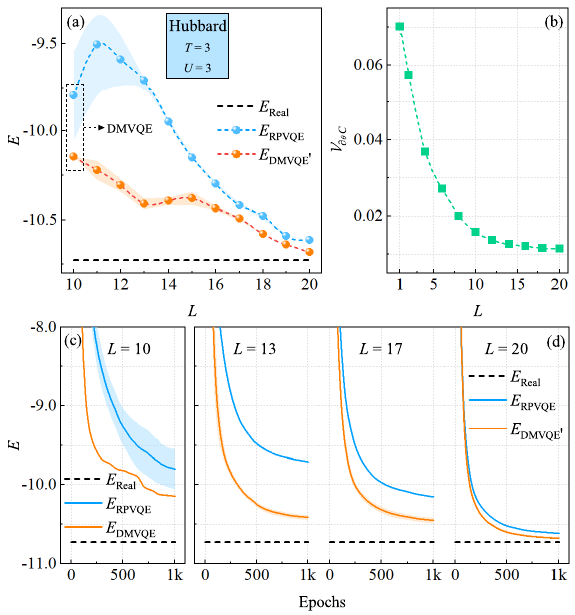}
  \caption{\textbf{Performance of DMVQE$'$ for Hubbard model.} (a) The ground-state energies ($E$) of both RPVQE and DMVQE$'$ as a function of $L$ for Hubbard model with $T=3$ and $U=3$. The number of epochs is set to 1000, and error bars represent the variance of the results from 30 repeated experiments. (b) The curve of trainability (${\cal V}_{\partial_\theta C}$) as a function of $L$. (c)-(d) Learning curves of ground-state energies ($E$) during training for RPVQE, DMVQE, and DMVQE$'$.}
  \label{fig6}
\end{figure}

Accurately calculating the ground state energy of the Hubbard model is crucial for understanding phenomena such as magnetism, superconductivity, and quantum phase transitions. However, RPVQE with shallow circuits often suffers from insufficient accuracy. Taking the Hubbard model with $T=3$ and $U=3$ as an example, we first evaluated whether DMVQE demonstrates performance improvements over RPVQE.  To ensure optimal performance and reliability, the number of epochs was set to 1000, and each experiment was repeated 30 times to obtain the average results. As highlighted by the dashed rectangle in  Fig.~\ref{fig6}(a), although DMVQE shows significant improvement over RPVQE, its accuracy remains insufficient. A noticeable gap exists between $E_{\rm DMVQE}$ and $E_{\rm Real}$. This limitation arises from the insufficient expressibility of the PQC with $L=10$ \cite{HEA}. 

Generally, PQCs must have enough expressibility to effectively generate pure states that adequately represent the solution space \cite{SECA}. In principle, adding more quantum gates and trainable parameters by increasing $L$ of PQCs can enhance their expressibility \cite{2-34}. However, there is a trade-off between expressibility and trainability \cite{EXPvsBP}: as PQCs become more expressive, the variance in the cost gradient decreases, leading to a flatter optimization landscape. This significantly hinders VQAs performance due to the issue of BP, a challenge that has gained wide concern \cite{SECA, LGnetwork, NNVQE,EXPvsBP-28,7-27-ent,16-8,16-9,16-13,16-14,16-16}. To visually illustrate how the optimization landscape of PQC evolves with increasing $L$, we computed the trainability with $L$ ranging from 1 to 20. BP is usually quantified in terms of unitary t-designs \cite{2-design, ENTvsBP-36,ENTvsBP-37}, with trainability dynamics depending solely on the variance of the partial derivatives (${\cal V}_{\partial_\theta C}$) \cite{bp-original}, which can be expressed as,
\begin{align}
{\cal V}_{\partial_\theta C}=\langle(i\langle0\vert{U^{\dagger}_{\mathcal{R}}}\lbrack{V_{k}},U^{\dagger}_{\mathcal{L}}HU_{\mathcal{L}}\rbrack{U_{\mathcal{R}}}\vert{0}\rangle)^2\rangle,
\label{eq10}	
\end{align}
where $U_{\mathcal{L}}$ and $U_{\mathcal{R}}$ represent a bipartite cut of a PQC, $V_{k} = \bigotimes_{j=1}^{n}\sigma_{j}$ is a Hermitian operator. See Ref.~ \cite{bp-original} for detailed derivation and explanation of Eq.~(\ref{eq10}). As $L$ increases from 1 to 10, ${\cal V}_{\partial_\theta C}$ decreases rapidly. Beyond $L=10$, the rate of decay slows, and between $L=15$ and $L=20$, ${\cal V}_{\partial_\theta C}$ almost converges to a flat landscape, as shown in Fig.~\ref{fig6}(b). This indicates that the PQCs with more than 10 layers indeed have extremely low trainability.

To address this issue, we try to assign DM-generated parameters to only a portion of deeper PQCs. Specifically, DM-generated parameters are assigned to the first few layers of a deeper PQC (the first 10 layers in our study), while the remaining layers are randomly initialized. To distinguish this parameter configuration method from the previous approach, we refer to it as DMVQE$'$, as shown in Fig.~\ref{fig5}. Performance of DMVQE$'$ with $L$ ranging from 10 to 20 are evaluated. It should be pointed out that $L=10$ corresponds to the original DMVQE. For each $L$, we compute the ground-state energies obtained by DMVQE$'$ ($E_{\rm DMVQE'}$) and, for comparison, the corresponding $E_{\rm RPVQE}$. As shown in Fig.~\ref{fig6}(a), both $E_{\rm RPVQE}$ and $E_{\rm DMVQE'}$ exhibit a decreasing trend as $L$ increases from 10 to 20. The key distinction lies in the significant fluctuations observed in $E_{\rm RPVQE}$, reflecting the inherent instability of randomly initialized parameters for RPVQE. Notably, for all  $L$, $E_{\rm DMVQE'}$ consistently outperforms $E_{\rm RPVQE}$, demonstrating the effectiveness of $E_{\rm DMVQE'}$ in mitigating BP issues caused by increasing circuit depth. Additionally, we observe that the advantage of $E_{\rm DMVQE'}$ over $E_{\rm RPVQE}$ diminishes as $L$ increases, suggesting that there is an upper limit to the number of randomly initialized layers in DMVQE$'$ to maintain its superiority.

We also plotted the learning curves of $E_{\rm RPVQE}$ and $E_{\rm DMVQE'}$ ($E_{\rm DMVQE}$) during the training process for $L=10$, $L=13$, $L=17$, $L=20$ to illustrate the optimization rates and performance simultaneously. As shown in Fig.\ref{fig6}(c), $E_{\rm RPVQE}$ exhibits significant variance, further highlighting the inherent instability of RPVQE caused by random initialization. In contrast, DMVQE experiments with identical parameter initialization yielded zero variance for $E_{\rm DMVQE}$. More importantly, both the optimization rate and final performance of $E_{\rm DMVQE}$ and $E_{\rm DMVQE'}$ surpass those of $E_{\rm RPVQE}$, as shown in Fig.\ref{fig6}(c)-(d). These results indicate that our approach has excellent adaptability even for Hubbard models with Hamiltonian structures entirely different from those in the dataset. Moreover, DMVQE$'$ can enhance the optimization efficiency in the case of low trainability, exhibiting some BP mitigation effects.

\section{Conclusion}

In this paper, we propose to leverage denoising diffusion models (DMs) to tackle the optimization challenges encountered by VQAs. The DM is trained on a few data points in the Heisenberg model parameter space and then can be guided to generate high-performance PQC parameters for a series of Hamiltonians with similar local interaction structures. Numerical experiments validate its effectiveness in improving optimization efficiency and reducing quantum resource consumption with the Heisenberg, Ising, and Hubbard models. 

Results demonstrate that for Heisenberg models with parameters not included in the training dataset, DMVQE can generate states corresponding to the exact solutions for 69\% of the Hamiltonians. For the remaining Hamiltonians, it can also obtain a value with MRE of less than 13\% from the exact solutions. Even for previously unseen  Hamiltonians, DMVQE can generate an appropriate initial state to achieve rapid convergence. Results of the Ising model prove a faster convergence of DMVQE compared with RPVQE.  Only after 50 epochs, the MRE of $E_\mathrm{DMVQE}$ is 6.79\%, much better than that of $E_\mathrm{RPVQE}$ with 41.26\%. We also noticed that in the optimization process, 28.9\% RPVQE cases are trapped in local minima while only 1.6\% DMVQE cases suffer from the same dilemma. At last, the results of the Hubbard model indicate that for deep circuits with extremely low trainability, assigning DM-generated parameters to only a portion of its layers can help mitigate the effects of BP. 
       
  In summary, the advantages of our method are twofold: first, it can generate an appropriate initial state closer to the exact solution, thus mitigating the effects of BP and local minima simultaneously; second, it applies to a wide range of Hamiltonians of interests, avoiding massive repeated quantum measurements and classical iterations. Our approach provides an efficient paradigm for optimization and may advance the practical application of VQAs.

\section{Acknowledgement}

This work was supported by the National Natural Science Foundation of China under Grant No.61975005, the Beijing Academy of Quantum Information Science under Grants No.Y18G28, and the Fundamental Research Funds for the Central Universities.

\bibliography{DMVQE.bib}


\end{document}